\documentclass[12pt]{iopart}
\usepackage{graphics}
\expandafter\let\csname equation*\endcsname\relax
\expandafter\let\csname endequation*\endcsname\relax
\usepackage{amsmath}

\newcommand{\ket}[1]{\left| #1 \right>} 
\newcommand{\ketbra}[2]{\mbox{$|#1\rangle\langle #2|$}}

\begin{document}
\title[The multi-site variational polaron transformation]{A multi-site variational master equation approach to dissipative energy transfer}

\author{Felix A Pollock$^1$, Dara P S McCutcheon$^{2,3}$, Brendon W Lovett$^{4,5}$, Erik M Gauger$^{5,6}$ and Ahsan Nazir$^3$}

\address{$^1$ Atomic \& Laser Physics, Clarendon Laboratory, University of Oxford, Parks Road, Oxford OX1~3PU, UK}
\address{$^2$ Departamento de F\'isica, FCEyN, UBA, Ciudad Universitaria, Pabell\'on 1, 1428 Buenos Aires, Argentina}
\address{$^3$ Blackett Laboratory, Imperial College London, London SW7~2AZ, UK}
\address{$^4$ SUPA, Institute of Photonics and Quantum Sciences, Heriot-Watt University, Edinburgh EH14~4AS, UK}
\address{$^5$ Department of Materials, Oxford University, Oxford, OX1~3PH, UK}
\address{$^6$ Centre for Quantum Technologies, National University of Singapore, 3 Science Drive 2, Singapore 117543}
\ead{\mailto{Felix.Pollock@physics.ox.ac.uk}, \mailto{a.nazir@imperial.ac.uk}}

\begin{abstract}
Unitary transformations can allow one to study 
open quantum systems in situations for which standard, weak-coupling type approximations are not valid. We develop here an extension of the variational (polaron) transformation approach to open system dynamics, which applies to arbitrarily large exciton transport networks with local environments. After deriving a time-local master equation in the transformed frame, we go on to compare the population dynamics predicted using our technique with other established master equations. 
The variational frame dynamics are found to agree with both weak coupling and full polaron master equations in their respective regions of validity. In parameter regimes considered difficult for these methods, the dynamics predicted by our technique are found to interpolate between the two. The variational method thus gives insight, across a broad range of parameters, into the competition between coherent and incoherent processes in determining the dynamical behaviour of energy transfer networks.
\end{abstract}

\section{Introduction}

The theory of open 
systems is necessary to describe any quantum system in contact with an uncontrollable and non-negligible environment. In problems of energy transport one is often interested in a regime where the environment, which consists of a very large number of degrees of freedom, is highly influential. We are here concerned with the dynamics of electronic excitations across some discrete network of molecules, in which the environment can play a key role --- and this is exemplified by recent observations in various microscopic biological systems. Examples include: Molecular complexes involved in photosynthesis, such as the \emph{Fenna-Matthews-Olsen} (FMO) complex in green sulphur bacteria \cite{engel07,panitchayangkoona10,ChengFleming2009,IshizakiFleming2009_2,ChinHuelga2012,TronrudWen2009,RitschelRoden2011,OlbrichJansen2011,renger98,mohseni08,PlenioHuelga2008} and  the light harvesting photosystems in green plants \cite{ScholesFleming2005}; the electron transport chain in \emph{Respiratory Complex I} \cite{DornerGoold2012}; the donor-bridge-acceptor model of olfaction \cite{BrookesHartoutsiou2007} and certain models of magnetoreception in birds \cite{RitzAdem2000,GaugerRieper2011}. Similar energy transport models have also been used extensively outside of biology --- applications range from the dynamics of coupled quantum dots \cite{CrookerHollingsworth2002,nazir09,rozbicki08,gerardot05} in solid state physics, to those of impurities in lattice Bose-Einstein condensates (BECs) \cite{KleinBruderer2007}.

Several techniques have been developed over the years to calculate the explicit time domain dynamics of open quantum systems. Some are numerically exact, meaning that given sufficient computational resources, they will converge to the correct dynamics under some well-controlled approximations. Such techniques include the path integral \cite{MakriMakarov1995_1,MakriMakarov1995_2,nalbach11}, hierarchical equations of motion (HEOM) \cite{IshizakiFleming2009_2,IshizakiFleming2009_1,TanakaTanimura2010} and density matrix renormalization group (DMRG) \cite{PriorChin2010} methods. 
Though powerful, these approaches typically 
place restrictions on the kind of system that 
can be modelled, and they may also 
scale badly (in terms of computing resources) with the size and complexity of said system \cite{IshizakiFleming2009_2,MakriMakarov1995_2}. Often --- as will be the case in this article --- more numerically tractable, though approximate, methods are used. For example, those based on master equations
\cite{BreuerPetruccione,RodenEisfeld2009,BerkelbachReichman2012}. This technique 
provides an equation of motion for the reduced density matrix of the system in question without having to track the full evolution of the environment, though normally involves some kind of perturbative expansion in a small parameter, such as the system-environment coupling strength. 
A further advantage of the master equation approach is that it can offer insights into the mechanisms underlying the dynamics of a system by relating rates and energy shifts directly to microscopic parameters. However, the obvious drawback of many master equations 
is that they rely on certain Hamiltonian parameters 
being small. 
If this condition is not fulfilled, then the truncation of the perturbative expansion 
often leads to (potentially unphysical) results which can diverge wildly from the true dynamics \cite{FlemingCummings2011}.

In certain parameter regimes, performing unitary transformations, such as the polaron transformation \cite{Holstein1959_1,Holstein1959_2,JacksonSilbey1983,nazir09,McCutcheonNazir2010,mccutcheon11,JangCheng2008,jang09,Jang2011,KolliNazir2011}, on the combined system-environment Hamiltonian can result in a smaller interaction energy in the transformed frame. The transformed system is then amenable to being modelled using a perturbative master equation. For example, the polaron transformation can work well over a broad range of parameters when the relevant environmental timescales 
are short compared to those in the system --- in fact, the polaron transformation diagonalizes the Hamiltonian we use below when 
no electronic couplings are present between the sites. It is thus often used when the coupling between system and environment is strong or when internal system couplings are small. Between the weak-coupling and polaron regimes, however, lies a region of parameter space for which neither model is appropriate. In addition, the polaron transformation runs into problems when applied to a system with an Ohmic or sub-Ohmic environment (one for which 
the environment spectral density scales linearly or sub-linearly, 
respectively, at low frequencies). In this case, 
infrared divergences arise which prevent certain master equations (such as the time-local form 
used in this work) from correctly predicting the dynamics in the transformed frame.

As an extension of the standard polaron approach, the variational (polaron) transformation \cite{SilbeyHarris1984,HarrisSilbey1985,McCutcheonDattani2011,McCutcheonNazir2011,NazirMcCutcheon2012,zimanyiRS,LeeMoix2012} allows one to attempt to derive a perturbative series which 
is as valid as possible (given the restricted form of the transformation) in all parameter regimes. This is achieved by performing an optimized, partial polaron 
displacement on each of the environmental phonon modes relevant to their particular mode frequency, thereby interpolating between the weak-coupling and polaron 
representations for separate modes, as well as in the final master equation. Here we build on previous work on the variational transformation for two-site systems \cite{McCutcheonDattani2011,McCutcheonNazir2011,zimanyiRS} which is in turn based on an idea originating with Silbey \& Harris \cite{SilbeyHarris1984,HarrisSilbey1985}. The major new contribution of this paper is the generalization of the formalism to any number of sites, allowing for the 
simulation of large networks across a range of 
environmental coupling parameters and temperatures. For comparison to other techniques, we have included examples of dynamics 
for systems in several different 
regimes.

In section \ref{sec:model} 
we describe the mathematical model for which the transformation is valid, including some of its limitations. We then go on to discuss the form of the variational polaron transformation and the accompanying optimization procedure in section \ref{sec:polaron}. In section \ref{sec:master} we outline the master equation formalism in the variationally transformed frame, and in section \ref{sec:results} we present some example dynamics, including that of the FMO system. Finally, in section \ref{sec:discussion}, we shall conclude by briefly discussing the various advantages and drawbacks of the method outlined herein.

\section{The transport model}
\label{sec:model}
\begin{figure}
\makebox[\textwidth][c]{\includegraphics{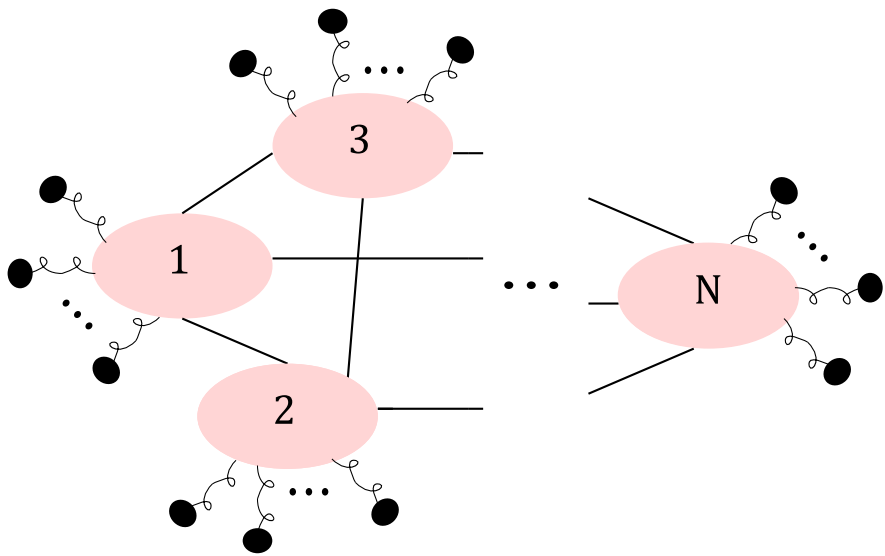}}
\caption{A schematic representation of the exciton transport network. The numbered ellipses represent the sites of the system, each of which is coupled to its own environment of phonon modes (masses on springs) and, in general, to every other site --- via the black lines.}
\label{nsite}
\end{figure}

The system ($S$) considered in this work is that of $N$ coupled two-level systems, known as sites. Between them they carry exactly one excitation --- for molecular networks these are electronic excitations which, for charge neutral systems, are called excitons. The latter restriction to a single excitation allows for a great reduction in the size of the system Hilbert space, and is sufficient to describe the behaviour of many physically and biologically relevant systems \cite{nazir09,rozbicki08,StaceDoherty2005}. For example, it is thought to be a valid approximation to the \emph{in vivo} dynamics of the FMO complex studied in section \ref{sec:results} \cite{ChinHuelga2012}. Each of the $N$ sites is linearly-coupled to its own, independent phonon environment ($E$). 
The Hamiltonian for the combined system and environment is given by
\begin{eqnarray}
H = H_{S} + H_{E} + H_{I}, \qquad	H_S = \sum_{n}{\varepsilon_n \ketbra{n}{n}} + \sum_{n \neq m}{V_{nm} \ketbra{n}{m}}, \nonumber \\
H_E = \sum_{n,\mathbf{k}}{ \omega_{n,\mathbf{k}} b^\dagger_{n,\mathbf{k}} b_{n,\mathbf{k}}}, \qquad
			H_I	= \sum_{n,\mathbf{k}}{\ketbra{n}{n}(g_{n,\mathbf{k}} b^\dagger_{n,\mathbf{k}} + g^*_{n,\mathbf{k}} b_{n,\mathbf{k}})}, \label{eq:ham}
\end{eqnarray}
where $b_{n,\mathbf{k}}$ is the annihilation operator for phonon mode $\mathbf{k}$ on site $n$, and $\ket{n}$ is the state of $S$ in which only site $n$ is excited (see figure \ref{nsite} for a cartoon visualization of the above Hamiltonian).

The environmental spectrum 
is usually taken to be a continuum, such that the couplings $g_{n,\mathbf{k}}$ can be 
described in terms of a spectral density $J_n(\omega)$. The spectral density is defined by
\begin{equation}
J_n(\omega) = \sum_{\mathbf{k}}{|g_{n,\mathbf{k}}|^2 \delta(\omega-\omega_k)};
\end{equation}
which takes into account the density of states, dispersion relation and interaction mechanism with the environment. In the continuum case, a good measure of the strength of the system-environment coupling at each site is the reorganization energy \cite{IshizakiFleming2009_1}:
\begin{equation}
\lambda_n = \int_0^\infty{\rmd\omega  \frac{J_n(\omega)}{\omega}}.
\end{equation}
This model ignores any spatial correlations between phonon excitations at different sites, meaning that 
the Hamiltonian in \ref{eq:ham} is not relevant for systems with strong, long-range correlations, such as impurities in BECs. For the case of FMO it has been claimed, based on detailed molecular dynamics simulations, that spatial correlations do not play a significant role in the exciton dynamics \cite{OlbrichStrumpfer2011}.  Since the environments at each site are independent, one can take the couplings to be real for this model without loss of generality. In the case of a global environment, however, the phases of the couplings to each site can encode the environmental correlations between them.

We assume that the environment is initially in a thermal (Gibbs) state at temperature $T = 1/(k_B \beta)$: $\rho_E(0) = \rme^{- \beta H_E} / \tr(\rme^{- \beta H_E})$. In addition, the combined system-environment state is assumed to be initially separable, such that there are no system-environment correlations: $\rho(0) = \rho_S(0) \otimes \rho_E(0)$.

\section{The variational polaron transformation}
\label{sec:polaron}
The Hamiltonian in \ref{eq:ham} describes the transport problem in a way which is intuitive and transparent, in that $H_S$ and $H_E$ are the Hamiltonians for the system and environment in isolation and the interaction term $H_I$ is simple in form. However, one of the features of quantum mechanics is that the physics of a system --- even a composite one --- is invariant under unitary transformations. As a consequence, 
one is not restricted to a single way of distinguishing two subsystems. In the model outlined above, much of the energy associated with the interaction between system and environment is due to the excitation deforming the surrounding molecular structure, and hence affecting the state of the phonon environment.

Applying the polaron transformation allows us to move into a reference frame where this back-action from system to environment is 
accounted for at the Hamiltonian level. The system is `dressed' by the environment, and the environmental phonon modes are displaced in phase space conditional on the state of the system. The explicit form of the transformation is
\begin{eqnarray}
\tilde{H} = \rme^G H \rme^{-G}, \quad {\rm where} \quad G = \sum_{n,\mathbf{k}}{\ketbra{n}{n} \omega_{n,\mathbf{k}}^{-1} (f_{n,\mathbf{k}} b^\dagger_{n,\mathbf{k}} - f^*_{n,\mathbf{k}} b_{n,\mathbf{k}})}, \label{eq:pol}
\end{eqnarray}
leading to a transformed Hamiltonian:
\begin{eqnarray}
\tilde{H} 			=   \tilde{H}_0 + \tilde{H}_I, \nonumber \\
					 			\tilde{H}_0 = \tilde{H}_{S} + \tilde{H}_{E},  \quad \tilde{H}_I = \tilde{H}_{L} + \tilde{H}_{D},  \\
\tilde{H}_{S}	= \sum_{n}{(\varepsilon_n + R_n) \ketbra{n}{n}} + \sum_{n \neq m}{\mathcal{B}_{n}\mathcal{B}_{m} V_{nm} \ketbra{n}{m}},  \nonumber \\
\tilde{H}_{L}	= \sum_{n,\mathbf{k}}{\ketbra{n}{n}  \left[(g_{n,\mathbf{k}}-f_{n,\mathbf{k}}) b^\dagger_{n,\mathbf{k}} + (g_{n,\mathbf{k}}-f_{n,\mathbf{k}})^* b_{n,\mathbf{k}}\right]}, \nonumber \\
\tilde{H}_{D}= \sum_{n \neq m}{ V_{nm} \ketbra{n}{m} B_{nm}}, \quad \tilde{H}_E = H_E .
\end{eqnarray}
The interaction Hamiltonian $\tilde{H}_I$ now contains two terms. One, $\tilde{H}_L$, is of the same linear form as the interaction in the untransformed Hamiltonian, albeit with modified coupling strength. The other term, $\tilde{H}_D$, contains a new kind of interaction between off-diagonal system operators 
and the environmental displacement 
operators
\begin{eqnarray}
\fl B_{nm} = B_{mn}^\dagger = \exp\left[\sum_{\mathbf{k}}{\omega_{n,\mathbf{k}}^{-1} (f_{n,\mathbf{k}} b^\dagger_{n,\mathbf{k}} - f^*_{n,\mathbf{k}} b_{n,\mathbf{k}})} - \sum_{\mathbf{k}}{\omega_{m,\mathbf{k}}^{-1} (f_{m,\mathbf{k}} b^\dagger_{m,\mathbf{k}} - f^*_{m,\mathbf{k}} b_{m,\mathbf{k}})}\right] \nonumber \\ - \mathcal{B}_{n}\mathcal{B}_{m} .  \nonumber
\end{eqnarray}
Here, the expectation values ($\mathcal{B}_{n}\mathcal{B}_{m}$) of the displacement operators in the $B_{nm}$ have been taken into the system Hamiltonian, and are thus treated as renormalized couplings between the sites. The $\mathcal{B}_n$'s are given by
\begin{eqnarray}
\mathcal{B}_n &= \tr\left\{ \exp\left[\sum_{\mathbf{k}}{\omega_{n,\mathbf{k}}^{-1} (f_{n,\mathbf{k}} b^\dagger_{n,\mathbf{k}} - f^*_{n,\mathbf{k}} b_{n,\mathbf{k}})}\right] \rho_E \right\} \nonumber \\
&= \exp\left[-\frac{1}{2} \sum_{\mathbf{k}}{\frac{|f_{n,\mathbf{k}}|^2}{\omega_{n,\mathbf{k}}^{2}} \coth(\beta \omega_{n,\mathbf{k}}/2)}\right], \nonumber
\end{eqnarray}
for a thermal equilibrium environmental state $\rho_E$.
The site energies after transformation are also shifted in comparison 
to the original frame by a factor $R_n$, defined as
\begin{eqnarray}
R_n = \sum_{\mathbf{k}}{\omega_{n,\mathbf{k}}^{-1}\left[|f_{n,\mathbf{k}}|^2-2\Re(f_{n,\mathbf{k}} g_{n,\mathbf{k}}^*)\right]}. \label{eq:rpars}
\end{eqnarray}

Usually when the polaron transformation is discussed in the literature, what is meant is the \emph{fully-displaced} version of \ref{eq:pol} \cite{Jang2011,KolliNazir2011} where $f_{n,\mathbf{k}} = g_{n,\mathbf{k}}$, for all $n$ and $\mathbf{k}$. This results in $\tilde{H}_L = 0$, which leaves only the new displacement interaction term in the Hamiltonian: $\tilde{H}_I = \tilde{H}_D$. In the variational case, however, the $f_{n,\mathbf{k}}$ are left as free parameters, and minimization over an upper bound on the free energy, as described below, 
determines their values.
The idea is that the optimization inherent to 
the variational approach allows us 
to minimize the effect of 
the interaction Hamiltonian $\tilde{H}_I$, given the transformation form. 
This is done here in order to validate the use of perturbation series in various master equation approaches, which must in practice be truncated at some finite order. In general, the `smaller' $\tilde{H}_I$, the more accurate the low-order dynamics are likely 
to be.

Since, most of the time, there is no single parameter in $\tilde{H}_I$ which determines exactly how small its effect is, we choose instead to optimize the variational transformation by minimizing the contribution of $\tilde{H}_I$ to the free energy (the average energy of a thermal state of the system). This choice is consistent with earlier variational treatments \cite{SilbeyHarris1984,HarrisSilbey1985} and ensures that the steady state of the resulting dynamics is as accurate as possible --- in equilibrium the free energy should be at a minimum. As it is generally impossible to find an exact analytical expression for the free energy, it is the Feynman-Bogoliubov upper bound \cite{BinneyDowrick} that we shall minimize. The bound is given by
\begin{equation}
A_B = -\frac{1}{\beta} \ln \left[ \tr(\rme^{-\beta \tilde{H}_0}) \right] + \langle \tilde{H}_I \rangle_{\tilde{H}_0} + \mathcal{O}\left(\langle \tilde{H}_I^2 \rangle_{\tilde{H}_0} \right),
\label{eq:FBbound}
\end{equation}
where $\langle X \rangle_{\tilde{H}_0} = \tr\left(X \rme^{-\beta \tilde{H}_0}\right)$. The true free energy $A$ is related to this bound by the inequality $A \leq A_B$. Given that we want to end up with $\tilde{H}_I$ small, it is reasonable to neglect the higher order terms in \ref{eq:FBbound} as a first approximation. Furthermore, the interaction Hamiltonian in the transformed frame has been constructed such that the second term goes to zero, $\langle \tilde{H}_I \rangle_{\tilde{H}_0} = 0$. Therefore, minimization amounts to maximizing the value of $\tr(\rme^{-\beta \tilde{H}_0})$. Although, perhaps counter-intuitively, $\tilde{H}_I$ now appears to be absent from $A_B$
its influence is, in fact, still present implicitly in $\tilde{H}_0$.

The transformed system Hamiltonian can be written as a function of the \emph{renormalization parameters} $\{ R_n, \mathcal{B}_n \}$, therefore the minimization condition can be written:
\begin{equation}
\frac{\partial A_B}{\partial f_{n,\mathbf{k}}} = \frac{\partial A_B}{\partial R_n}\frac{\partial R_n}{\partial f_{n,\mathbf{k}}} + \frac{\partial A_B}{\partial \mathcal{B}_n}\frac{\partial \mathcal{B}_n}{\partial f_{n,\mathbf{k}}} = 0 \qquad \forall \; n, \mathbf{k},
\end{equation}
which, after using the expressions for the renormalization parameters in \ref{eq:rpars} allows us to write $f_{n,\mathbf{k}} = F_n(\omega_{n,\mathbf{k}}) g_{n,\mathbf{k}}$, with
\begin{equation}
F_n(\omega_{n,\mathbf{k}},\{ R_n, \mathcal{B}_n \}) = \frac{2 \omega_{n,\mathbf{k}} \frac{\partial A_B}{\partial R_n}}{2 \omega_{n,\mathbf{k}} \frac{\partial A_B}{\partial R_n}- \mathcal{B}_n \frac{\partial A_B}{\partial \mathcal{B}_n} \coth(\beta \omega_{n,\mathbf{k}} / 2)}. \label{eq:mincon}
\end{equation}
In the continuum limit for the environment, the minimization procedure for an $N$-site system amounts to solving the $2N$ coupled integral equations given by the definitions of the renormalization parameters:
\begin{eqnarray}
\mathcal{B}_n = \exp\left[-\frac{1}{2} \int_0^\infty{\rmd\omega \; \frac{F_n(\omega,\{ R_n, \mathcal{B}_n \})^2 J_n(\omega)}{\omega^{2}} \coth(\beta \omega/2)}\right], \label{eq:simeq1}\\
R_n = \int_0^\infty{\rmd\omega \; \frac{J_n(\omega)}{\omega} F_n(\omega,\{ R_n, \mathcal{B}_n \})\left[ F_n(\omega,\{ R_n, \mathcal{B}_n \})-2 \right]}. \label{eq:simeq2}
\end{eqnarray}

\section{Master equation formulation in the variational frame}
\label{sec:master}
As outlined above, a truncated perturbative expansion in the new interaction Hamiltonian $\tilde{H}_I$ should, following the optimization procedure, be as accurate as possible, given the polaron form of the 
transformation and the minimization condition used. The next step, therefore, is to derive a master equation in the variational frame using standard techniques. By utilizing a projection operator $\mathcal{P}$ with the following action on the combined system-environment state: $\mathcal{P} \tilde{\rho} = \tr_E(\tilde{\rho}) \otimes \rho_R$, where $\rho_R$ is an arbitrary reference state for the environment, we can separate out the reduced system dynamics. Here, we choose to derive a time-local or `time-convolutionless' master equation due to the relative ease with which it can be solved numerically. In addition, we cut off the perturbation series at second order. The resulting master equation has the following general form in the interaction picture \cite{BreuerPetruccione}:
\begin{equation}
\frac{\partial}{\partial t}  \tr_E\left\{\mathcal{P} \tilde{\rho}(t)\right\} = \tr_E\left\{\mathcal{K}_2(t) \mathcal{P} \tilde{\rho}(t)\right\} + \tr_E\left\{\mathcal{I}_2(t) (1-\mathcal{P}) \tilde{\rho}(0)\right\},
\label{eq:tcl}
\end{equation}
where $\mathcal{K}_2$ and $\mathcal{I}_2$ are superoperators acting on $\mathcal{P} \tilde{\rho}$ and $(1-\mathcal{P}) \tilde{\rho}$ respectively, which have been curtailed to second order in $\tilde{H}_I$.

In the untransformed frame the separable initial state means that the second, inhomogeneous term in \ref{eq:tcl} disappears for the choice $\rho_R = \rho_E(0)$. In the variational frame this is no longer the case and the inhomogeneous term must be taken into account. However, for two-site systems the inhomogeneous term was seen to have only a small, transient effect on the dynamics at finite temperatures \cite{McCutcheonNazir2011} for single site initial excitations. Therefore, we shall henceforth neglect it even in the transformed frame. This amounts to assuming that the environment relaxes into its displaced state instantaneously. One would expect this to be a good approximation at finite temperatures for smooth spectral densities and when the typical environment timescales are shorter than the transition timescales in the system \cite{JangCheng2008,jang09,Jang2011,KolliNazir2011,McCutcheonNazir2011}. The examples we present in section \ref{sec:results} satisfy each of these conditions.

The remaining (homogeneous) term is written explicitly as
\begin{equation}
\tr_E\left\{\mathcal{K}_2(t) \mathcal{P} \tilde{\rho}(t)\right\} = - \int_0^t{\rmd s \; \tr_E\left\{\left[\tilde{H}_I(t),[\tilde{H}_I(s),\mathcal{P} \tilde{\rho}(t)]\right]\right\}}.
\end{equation}
By writing the interaction Hamiltonian in the form $\tilde{H}_I = \sum_{i=1}^{N^2}S_i \otimes E_i$ (with interaction picture counterpart $\tilde{H}_I(t) = \sum_{i=1}^{N^2}S_i(t) \otimes E_i(t)$ ) we can rewrite the master equation in terms of system operators $S_i$ and two-time environmental correlation functions $\Lambda_{ij}(t-s) = \tr_E\left\{E_i(t)E_j(s) \rho_R\right\}$. After moving back into the Schr\"odinger picture, the master equation takes the form:\\\\
\begin{eqnarray}
\fl\frac{\partial \tilde{\rho}_S(t)}{\partial t} = - \rmi [\tilde{H}_S,\tilde{\rho}_S(t)] \nonumber \\
- \sum_{i,j} \int_0^t{\rmd s \; \left(\Lambda_{ij}(s)\left\{S_i S_j(s) \tilde{\rho}_S(t) - S_j(s) \tilde{\rho}_S(t) S_i\right\} + {\rm h.c.}\right)}.
\label{eq:me}
\end{eqnarray}
The interaction Hamiltonian system operators 
can be split into three distinct groups in the following way:
\begin{eqnarray}
S_i = \left\{
\begin{array}{lrclr}
\ketbra{n}{n} = S^z_n \quad &1 \leq&i&\leq N,&\\
\ketbra{n}{m} + \ketbra{m}{n} = S^x_{nm} \quad &N <&i&\leq \frac{1}{2}N(N+1),&\\
\rmi\ketbra{n}{m} - \rmi\ketbra{m}{n} = S^y_{nm} \quad  &\frac{1}{2}N(N+1) <&i&\leq N^2,&
\end{array} \right.
\end{eqnarray}
which leads in turn to three varieties of non-zero time correlation function. The first type are due to the linear interaction term, 
$\tilde{H}_L$, and are therefore of the same form as those that appear in the standard weak coupling master equation:
\begin{eqnarray}
\Lambda^{zz}_{n}(t) &= \phi^{zz}_n(t) \nonumber \\
&= \int_0^\infty{\rmd\omega \; J_n(\omega)\left[1-F_n(\omega)\right]^2\left[\cos(\omega t) \coth(\beta \omega /2) - \rmi \sin(\omega t)\right]}, \label{eq:phizz}
\end{eqnarray}
where $F_n(\omega)$ is the continuum version of the optimized function in \ref{eq:mincon}. The second type come from the displacement operator interaction, $\tilde{H}_D$, and are the only type to appear in the fully displaced polaron master equation:
\begin{eqnarray}
\fl\Lambda^{xx}_{nmpq}(t) = \frac{1}{2}V_{nm}V_{pq}\mathcal{B}_n \mathcal{B}_m \mathcal{B}_p \mathcal{B}_q  \bigg\{ \exp&\left[\delta_{np} \phi^{xy}_n(t) +\delta_{mq} \phi^{xy}_m(t) \right] \nonumber \\
&+ \exp \left[-\delta_{np} \phi^{xy}_n(t) -\delta_{mq} \phi^{xy}_m(t) \right]-2\bigg\}, \nonumber
\end{eqnarray}
\begin{eqnarray}
\fl\Lambda^{yy}_{nmpq}(t) = \frac{1}{2}V_{nm}V_{pq}\mathcal{B}_n \mathcal{B}_m \mathcal{B}_p \mathcal{B}_q \bigg\{ \exp&\left[\delta_{np} \phi^{xy}_n(t) +(\delta_{mq}-\delta_{mp}) \phi^{xy}_m(t) \right] \nonumber \\
&- \exp\left[-\delta_{np}\phi^{xy}_n(t) -(\delta_{mq}-\delta_{mp})\phi^{xy}_m(t) \right]\bigg\},
\end{eqnarray}
where
\begin{eqnarray}
\phi^{xy}_n(t) = \int_0^\infty{\rmd\omega \; \frac{J_n(\omega)}{\omega^2} F_n(\omega)^2 \left[\cos(\omega t) \coth(\beta \omega /2) - \rmi \sin(\omega t)\right]},\label{eq:phixy}
\end{eqnarray}
and the $\delta_{nm}$ are Kronecker deltas. Finally, the third type appear in the more general variational master equation due to an overlap between the two types of interaction:
\begin{equation}
\Lambda^{yz}_{nmp}(t) = \delta_{np} V_{nm}\mathcal{B}_n \mathcal{B}_m \phi^{yz}_n(t),
\end{equation}
with
\begin{eqnarray}
\phi^{yz}_n(t) = \int_0^\infty\rmd\omega \; \frac{J_n(\omega)}{\omega} F_n(\omega)\left[1-F_n(\omega)\right] \left[\sin(\omega t) \coth(\beta \omega /2) + \rmi \cos(\omega t)\right].\nonumber\\\label{eq:phiyz}
\end{eqnarray}

The dynamics calculated using \ref{eq:me} will be of the density matrix in the variationally transformed frame. In order to consider quantities in the original frame, one must perform the inverse of the transformation in \ref{eq:pol}. The site populations (diagonal elements of the density matrix) are unchanged, since the operators $\ketbra{n}{n}$ commute with the transformation. However, in general, the inverse transformations of the coherences (off-diagonal elements of the density matrix) are much more difficult to calculate. In the case that the inhomogeneous term in \ref{eq:me} is ignored, one can make the approximation \cite{KolliNazir2011} $(\rho_S(t))_{nm} = \mathcal{B}_n\mathcal{B}_m (\tilde{\rho}_S(t))_{nm}$ for $n\neq m$, where $\tilde{\rho}_S(s)$ is the system density matrix in the variational frame and $\rho_S(s)$ is the system density matrix in the untransformed frame. This is equivalent to making a Born approximation in the transformed frame. Under such an approximation the transformed frame state is $\tilde{\rho}(t)\simeq \tilde{\rho}_S(t)\otimes \rho_R$ for all times. This state will transform in the same way under the inverse variational transformation as the system Hamiltonian \ref{eq:ham} does under the forward transformation, leading to the factors of $\mathcal{B}_n\mathcal{B}_m$ for the off-diagonal elements mentioned above.

\section{Example system dynamics}
\label{sec:results}
\subsection{Three sites}
\begin{figure}
\makebox[\textwidth][c]{\includegraphics{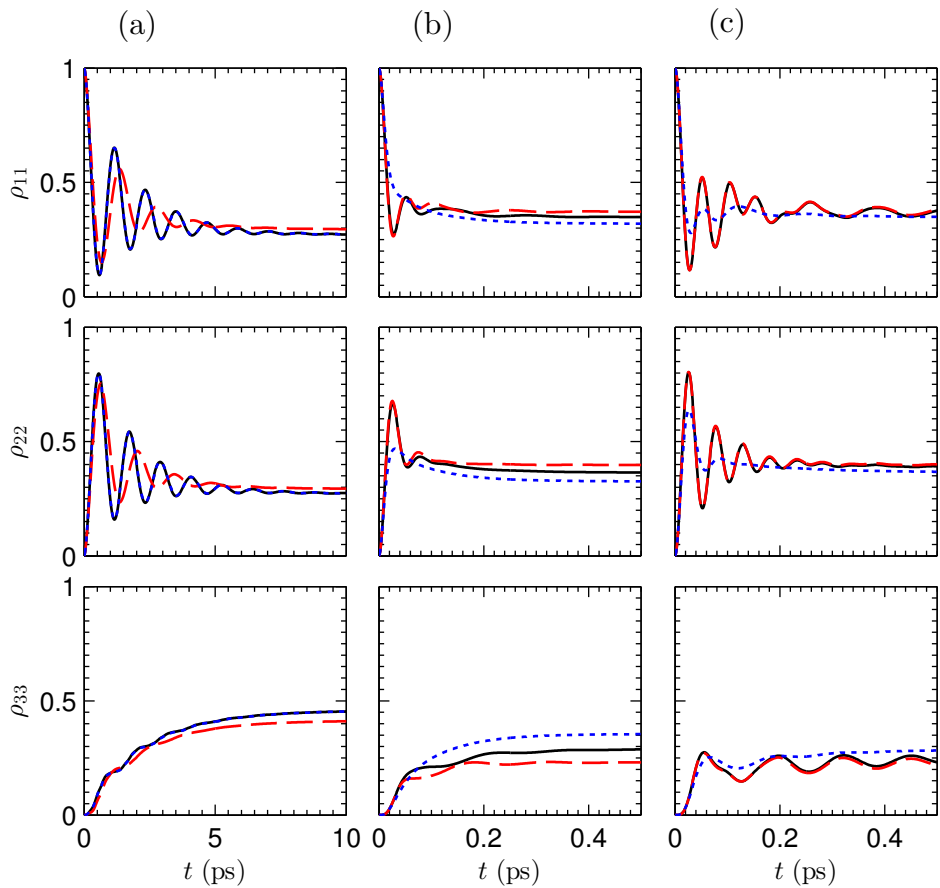}}
\caption{Population dynamics of a three-site system calculated using the variational (solid black), full-polaron (blue dotted) and weak-coupling (red dashed) master equations plotted as a function of time (in ps). The system parameters and spectral densities are as described in the text and in \ref{eq:specden}, with $\omega_c = 200 {\rm cm}^{-1}$, $E_1=E_2=50{\rm cm}^{-1}$, $E_3=0{\rm cm}^{-1}$ and $T=300\,{\rm K}$ in all cases. The remaining parameters for the individual subplots are: (a) $\lambda_1 = \lambda_2 = 60{\rm cm}^{-1}$, $\lambda_3 = 120{\rm cm}^{-1}$, $V_{12}=V_{23}=20{\rm cm}^{-1}$; (b) $\lambda_1=\lambda_2=\lambda_3 = 180{\rm cm}^{-1}$, $V_{12}=300{\rm cm}^{-1}$, $V_{23}=100{\rm cm}^{-1}$; (c) $\lambda_1=\lambda_2=\lambda_3 = 60{\rm cm}^{-1}$, $V_{12}=300{\rm cm}^{-1}$, $V_{23}=100{\rm cm}^{-1}$.}
\label{threesite}
\end{figure}

We would like to compare dynamics calculated in the variational frame, using \ref{eq:me}, with other techniques. After two sites, the next simplest system with a Hamiltonian of the form of that in \ref{eq:ham} has three sites and only nearest neighbour couplings. Figure \ref{threesite} shows the dynamics for such a system in a variety of parameter regimes, calculated in the variational frame. For comparison, we have also plotted the dynamics calculated in the fully-displaced polaron frame as well as that calculated using the untransformed Hamiltonian (weak-coupling, or Redfield, approximation). The system is characterized by its on-site energies $\{E_n\}$, inter-site couplings $V_{12}$ and $V_{23}$, and spectral densities of the form
\begin{equation}
J_n(\omega) = \frac{\lambda_n}{2}\frac{\omega^3}{\omega_c^3}{\rm e}^{-\omega/\omega_c}. \label{eq:specden}
\end{equation}
Column (a) in figure \ref{threesite} represents a regime where system frequencies ($\sim20{\rm cm}^{-1}$) are much smaller than the environment cutoff frequency ($200{\rm cm}^{-1}$). In this case, the fully-displaced polaron transformation is expected to do well \cite{NazirMcCutcheon2012}, and we would also expect the variational transformation to match it, as it indeed does. The untransformed dynamics fail to reach the correct steady state due to a reasonably large reorganization energy for the environment. We also found that the weak coupling approximation can lead to unphysical results for parameters differing from those in column (a) only by their intersite coupling. This was not the case for the dynamics in the variational or full polaron frames.

The second column, (b), shows dynamics in a regime for which neither the weak-coupling nor full-polaron transformation are ideally suited. System frequencies are comparable to environment frequencies, and the coupling to the environment is not small. The variational dynamics appear to interpolate between the two other results. It is clear that it agrees with the weak-coupling dynamics at short times, before settling on a different set of long time populations. The comparatively large system energies prevent the polaron transformation from dealing correctly with the lower frequency parts of the environment, leading to the dynamics in the full-polaron frame overestimating the damping of coherent oscillations. The variational transformation
preserves coherence here precisely due to the fact that 
it 
optimizes the frequency dependence of the polaron transformation, as opposed to indiscriminately displacing every phonon mode by the full amount.

Finally, in column (c) the reorganization energy is much lower than in column (b), but the system frequencies are still large. A weak coupling approximation is therefore valid in this case, and one would expect the dynamics in the untransformed frame to be more accurate than that in the (full) polaron frame. The red and black curves in this panel almost sit on top of each other, showing that the variational dynamics agree with the weak coupling results, and corroborates the fact that the variational transformation allows us to capture the dynamics across a broad range of coupling strengths. Note that, interestingly, both the weak-coupling and polaron approaches can overestimate the damping of coherence in comparison to the variational method, dependent on the parameter regime (cf. panels (a) and (c)).

\begin{figure}
\makebox[\textwidth][c]{\scalebox{0.87}{\includegraphics{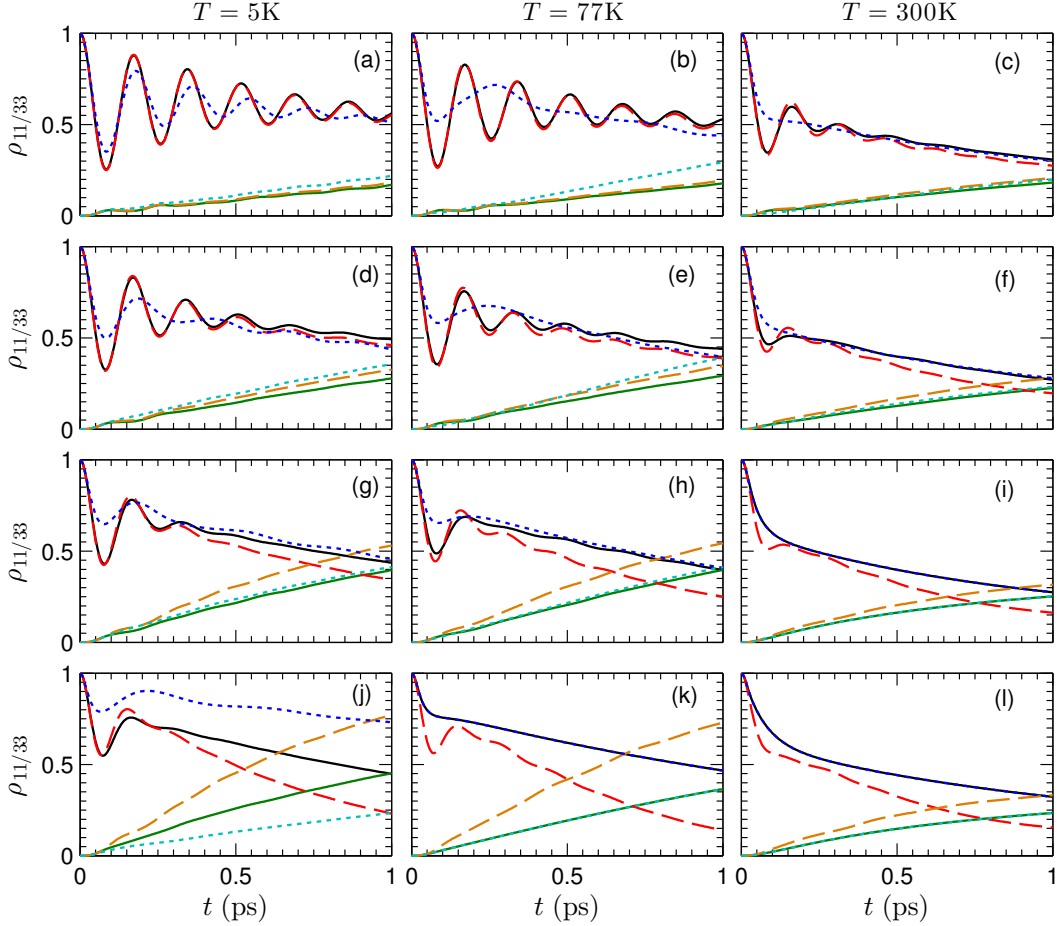}}}
\caption{Variational (solid), weak-coupling (dashed) and full-polaron (dotted) dynamics for the populations of site 1 (black, red, blue) and site 3 (olive, orange, cyan) of the FMO complex. Although all seven sites were modelled, only the input (1) and output (3) site populations are shown for clarity. The system Hamiltonian used is the same as that in \cite{NalbachBraun2011} and the spectral density is given in \ref{eq:nalbspec}. Panels (a-c) have $\eta = \frac{1}{2}$, (d-f) have $\eta = 1$, (g-i) have $\eta = 2$, and (j-l) have $\eta = 4$. The dynamics in the first column of plots were calculated at $5\, \rm K$, the second a $77\, \rm K$, and the third at $300\, \rm K$.}
\label{sevensite}
\end{figure}

\subsection{The Fenna-Matthews-Olsen complex}
We now analyse how the variational master equation performs for a larger system, namely the 
FMO complex. This is usually assumed to be a seven site system (although recent results suggest there is in fact an additional eighth site \cite{OlbrichJansen2011}) and thus has 
a much larger parameter space than the three site system in figure \ref{threesite}. Despite this, one can see from figure \ref{sevensite}, which compares FMO dynamics across a range of reorganization energies and temperatures, that the variational transformation performs the same kind of interpolation between weak-coupling and full-polaron dynamics as in the three-site case. The FMO system Hamiltonian used in this section --- taken from \cite{NalbachBraun2011} --- is:
\begin{equation}
H_{FMO} = 
\begin{pmatrix}
240 & -87.7 & 5.5 & -5.9 & 6.7 & -13.7 & -9.9 \\
-87.7 & 315 & 30.8 & 8.2 & 0.7 & 11.8 & 4.3 \\
5.5 & 30.8 & 0 & -53.5 & -2.2 & -9.6 & 6.0 \\
-5.9 & 8.2 & -53.5 & 130 & -70.7 & -17.0 & -63.3 \\
6.7 & 0.7 & -2.2 & -70.7 & 285 & 81.1 & -1.3\\
-13.7 & 11.8 & -9.6 & -17.0 & 81.1 & 435 & 39.7\\
-9.9 & 4.3 & 6.0 & -63.3 & -1.3 & 39.7 & 245
\end{pmatrix} 
{\rm cm}^{-1},
\end{equation}
and the spectral density --- the smooth part of that from \cite{NalbachBraun2011} --- is of the form:
\begin{equation}
J_n(\omega) = 3.053 \times 10^{-5} \times \eta\frac{\omega^5}{\omega_1^4}{\rm e}^{-\sqrt{\omega/\omega_1}}+1.908 \times 10^{-5} \times \eta\frac{\omega^5}{\omega_2^4}{\rm e}^{-\sqrt{\omega/\omega_2}}, \label{eq:nalbspec}
\end{equation}
with $\omega_1 = 0.575{\rm cm}^{-1}$ and $\omega_2 = 2{\rm cm}^{-1}$.

Figure \ref{sevensite} provides a clear example of the importance of the interplay between coherent and incoherent dynamics in excitonic energy transport. The biological purpose of the FMO complex is to transport excitations from site 1 (or sometimes site 6) to site 3, from which the excitation is then removed \cite{ChengFleming2009}. It is therefore beneficial to have the population on site 3 build up as fast as possible. One can see from the figure (most clearly in the second column), that the optimum rate of transfer (panels (j), (h) and (i)) occurs in the variational theory when coupling to the environment is neither too strong nor too weak. That is, phonon-assisted transport is enhanced by the presence of some degree of coherence in the system. These optimal cases appear to lie in the intermediate region of parameter space, outside the remit of weak-coupling or polaron master equations, for which something like the variational approach is required.

\begin{figure}
\makebox[\textwidth][c]{\scalebox{0.87}{\includegraphics{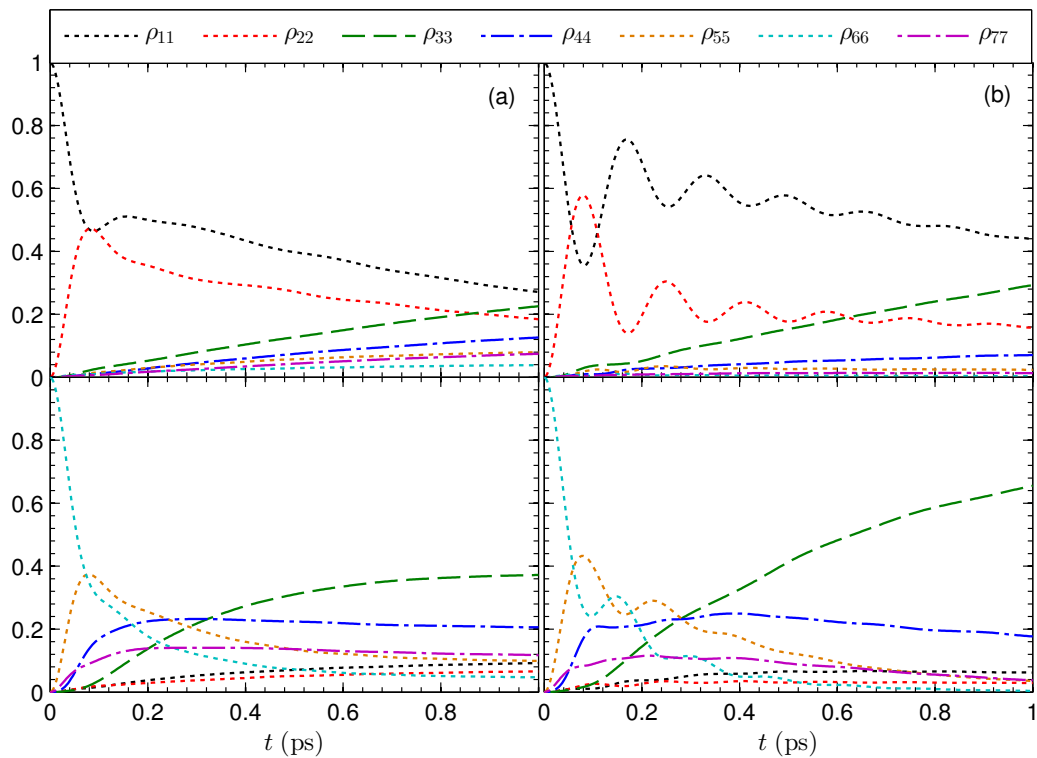}}}
\caption{Variational frame population dynamics for the FMO complex with the same parameters as figure 2 of \cite{NalbachBraun2011}, with the exception that only the smooth, non-peaked part of the spectral density was used. The dynamics in panel (a) were calculated at $300\, \rm K$ whilst those in (b) were calculated at $77\, \rm K$. Two different initial states: $\rho_S(0) = \ketbra{1}{1}$ and $\rho_S(0) = \ketbra{6}{6}$ were used for the upper and lower plots, respectively.}
\label{specific}
\end{figure}

The full, seven-site dynamics for the cases where $\eta = 1$, at $T = 77\, {\rm K}$ and $T = 300\, {\rm K}$, are shown in figure \ref{specific}. These plots correspond to the parameter regimes of the FMO dynamics in figure 2 of \cite{NalbachBraun2011}. However, the exact calculations presented in that paper include a significant peak in the spectral density which, in any master equation approach, would ideally be treated separately from the rest of the environment in order to capture its effect on the dynamics non-perturbatively. Moreover, a variational polaron treatment of such a peak would likely cause there to be significant system-environment correlations in the transformed frame, which would not be taken into account without the inclusion of inhomogeneous terms in the master equation. That being said, the qualitative agreement of the variational dynamics shown in the figure with the results in \cite{NalbachBraun2011} is surprisingly good.

\section{Discussion}
\label{sec:discussion}
There are several advantages to using master equations over other approaches to open quantum systems dynamics. Primarily, these are the efficiency with which one can solve them, and the potential insights into underlying physics which they can give. 
We can see from the various terms in \ref{eq:me} exactly how the parameters in the Hamiltonian enter into the system dynamics and the relative magnitude of these terms can give us an idea of the parameter regime of a given system --- in the sense of which quantities are contributing most to the dynamics.

Moving into the variational frame prior to solving the master equation allows us 
to calculate sensible dynamics over a larger range of parameters compared to the more standard weak-coupling or full-polaron approaches. As can be seen from section \ref{sec:results}, the variational master equation can capture the dynamics in both the weak and strong coupling regimes, when the weak coupling and polaron master equations, respectively, are expected to do well. It is also able to bridge the gap between the two in intermediate regimes.

The variational master equation is expected to work well over a wide range of parameters, and can in principle handle arbitrary spectral densities. However, like the full polaron transformation, it works best in the scaling limit \cite{NazirMcCutcheon2012}, in which important environmental frequencies ($\omega_c$) are large compared to relevant system frequencies ($V_{nm}$). The corresponding downside is that moving into the variational frame provides less of an advantage in terms of 
improving the accuracy of the dynamics when the typical environmental timescales are significantly longer than those of the system, and the coupling between the two is strong. The intuition for this is that the low frequency phonon modes are too `sluggish' to keep up with the motion of the exciton as it moves through the system and do not, therefore, dress the system in the same way as higher frequency modes. They may still, however, have a profound impact on the system dynamics, which a transfomation of displacement form is unable to capture.

One aspect of our method which might benefit from modification is the specific minimization condition used. Whilst we expect the first term of \ref{eq:FBbound} to be a good metric for the size of the interaction, it does not directly correspond to the quantity which we expand perturbatively in the master equation. In fact, the next highest order term, ($\mathcal{O}(H_I^2)$), is mathematically more similar \cite{ChinTurlakov2006}, albeit much more complicated. As one smoothly varies the Hamiltonian parameters, the optimum transformation can jump, as different local minima become global minima. This effect has been studied for the case of two sites \cite{LeeMoix2012}, and it was found that the variational predictions 
are less accurate around the discontinuity. For multiple sites, the free energy landscape becomes more complex and more local minima emerge in parameter space, hence there is greater scope for this kind of jumping. Whilst not optimal, the transformations corresponding to such local minima are still likely to lead to more accurate dynamics than those calculated in the untransformed frame.

In summary, we have outlined a variational method for solving open quantum systems dynamics in molecular networks with local environments. The method is valid over a wide range of parameters and is efficient to compute. By moving into a reference frame in which system and environment are less strongly interacting, one is able to use a perturbative master equation to more accurately calculate dynamics. For our Hamiltonian \ref{eq:ham}, the approach surpasses both weak-coupling and full-polaron master equations in terms of breadth of applicability. It can be used to model interesting biological systems which sit in difficult intermediate coupling regimes, such as FMO, and allows for the systematic study of the effects of certain parameters on the dynamics.

There is still ample room for improvement, and the general concept of redrawing the boundary between system and environment has far greater reach than the implementation presented in this paper. For instance, one could generalize the transformation to a larger class of Hamiltonians, say those whose environments couple to multiple sites, or one could augment the form of the transformation itself, perhaps by including squeezing in addition to displacement. The technique we have developed utilizes one of the most fundamental properties of quantum mechanics, namely the invariance of dynamical laws under unitary transformations, and gives insight into the important physical mechanisms underlying the evolution of open quantum systems. Whilst future master equation approaches may go beyond the polaron transformation, they are likely to benefit from a kind of variational minimization in the spirit of that which we have outlined here.

\section{Acknowledgements}
We would like to thank the Leverhulme Trust and the National Research Foundation and Ministry of Education of Singapore for financial support. BWL thanks the Royal Society for a University Research Fellowship. DPSM thanks the EPSRC, CHIST-ERA project SSQN, and CONICET. AN is supported by Imperial College. In addition, we are grateful to Alex Chin, Agata Checinska, and Libby Heaney
for useful discussions.

\section*{References}
\bibliography{MultisiteVariational}
\bibliographystyle{iopart-num}

\end{document}